\documentclass{ws-ijmpd}
\usepackage[super,compress]{cite}
\usepackage{amsmath}
\usepackage{amssymb}
\usepackage{hyperref}
\newcommand{\dd}{\, {\rm d}}

\usepackage{cancel}
\usepackage{bigstrut}
\newcommand{\rs}{r_{\rm s}}

\usepackage{subfigure}

\newcommand{\eff}{_{\rm eff}}

\newcommand{\pn}{\Phi_{\rm N}}

\begin{document}

\markboth{J. Sakstein, B. Jain and V. Vikram}
 {Detecting Modified Gravity in the Stars}

%
%

\title{Testing Gravity Theories Using Stars}

\author{Jeremy Sakstein}

\address{Department of Applied Mathematics and Theoretical Physics,\\ Centre for
Mathematical Sciences, University of Cambridge,\\ Wilberforce Road, Cambridge
CB3 0WA, UK\\ and\\
Perimeter Institute for Theoretical Physics,\\ 31 Caroline St. N,
Waterloo, ON, N2L 6B9, Canada\\j.a.sakstein@damtp.cam.ac.uk}

\author{Bhuvnesh Jain}

\address{Department of Physics \& Astronomy, University of
Pennsylvania,\\ Philadelphia, PA 19104, USA\\
bjain@physics.upenn.edu}

\author{Vinu Vikram}
\address{Department of Physics \& Astronomy, University of
Pennsylvania,\\ Philadelphia, PA 19104, USA\\
vinu@physics.upenn.edu}
\maketitle

\begin{abstract}
Modified theories of gravity have received a renewed interest due to their
ability to account for the cosmic
acceleration. In order to satisfy the solar system tests of gravity, these
theories need to include a screening mechanism that
hides the modifications on small scales. One popular and well-studied theory is
chameleon gravity. Our own galaxy is necessarily
screened, but less dense dwarf galaxies may be unscreened and their constituent
stars can exhibit novel features. In particular,
unscreened stars are brighter, hotter and more ephemeral than screened stars in
our own galaxy. They also pulsate with a shorter
period. In this essay, we exploit these new features to constrain chameleon
gravity to levels three orders of magnitude lower the previous measurements.
These constraints are currently the strongest in the literature.
\end{abstract}




\section{Introduction}

Einstein's general relativity has been the cornerstone of gravitational physics
for nearly a century, but how well have we really
tested it? To date, experimental tests have been limited to our own solar system
and a handful of isolated
systems such as binary pulsars and it has passed each with flying colours. We
have not yet tested gravity in other galaxies or on
larger --- inter-galactic and cosmological --- scales. Indeed, when one allows
the theory of gravity to vary and examines
constraints
coming from linear cosmological probes, there are large regions in theory space
that are consistent with the current
observations \cite{Bean:2010zq}. The apparent acceleration of the universe
\cite{Riess:1998cb,Perlmutter:1998np,Copeland:2006wr}
is one of the biggest unsolved problems in modern physics and has led to a
recent interest in modified theories of gravity as one
possible explanation \cite{Clifton:2011jh}. This then presents a problem: how
can these modifications be large enough to give
interesting cosmological dynamics whilst still
satisfying the solar system bounds?

\begin{figure}[t,h]
\centering
\includegraphics[width=10cm]{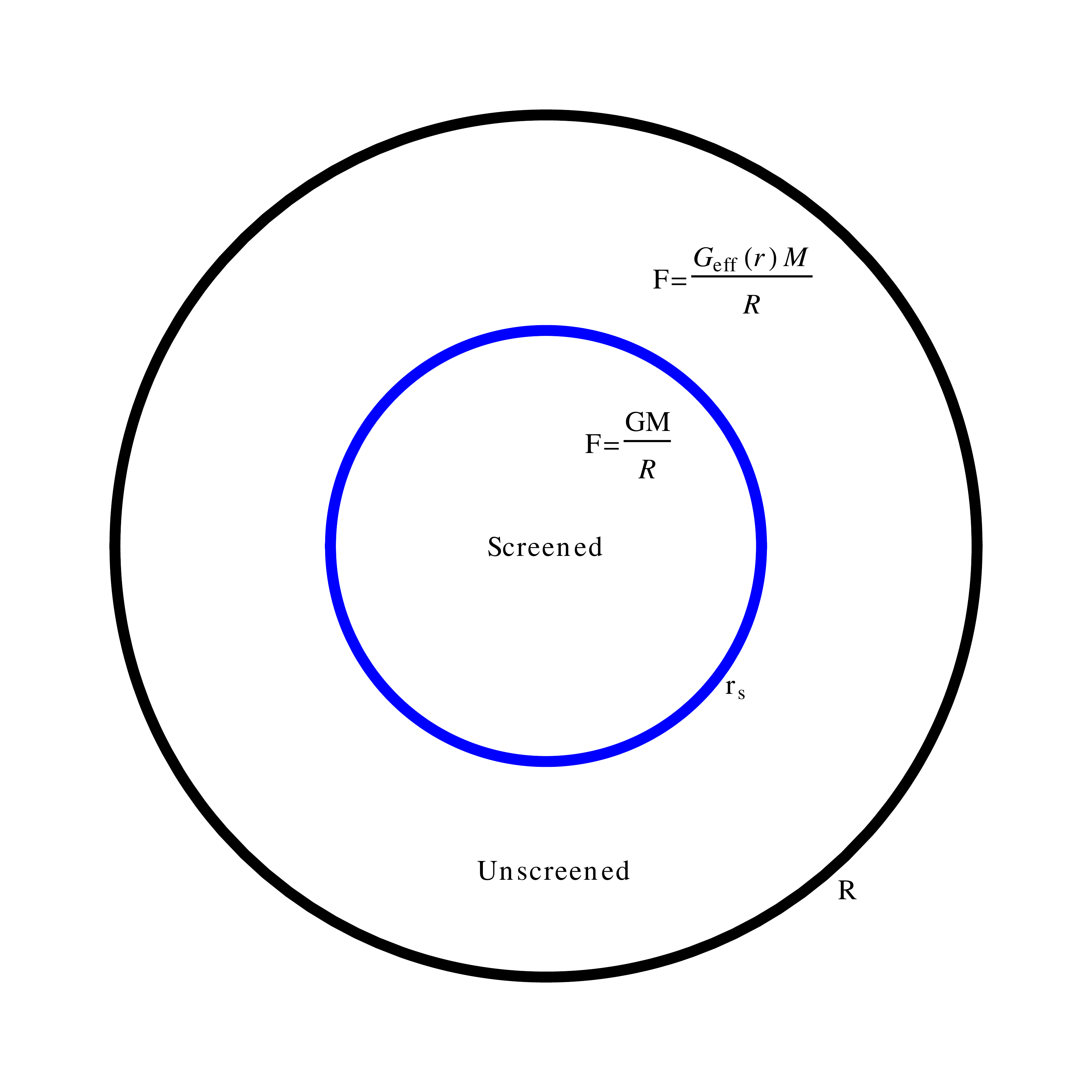}
\caption{The screening profile of a spherical object in chameleon gravity. The
region interior to the screening radius $\rs$ is 
screened and the force per unit mass (indicated by $F$) is simply the Newtonian
force. The exterior region is unscreened and
the force per unit mass is given by the Newtonian force law with $G\rightarrow
G\eff$.} 
\label{fig:screening}
\end{figure}

The solution is to construct theories of gravity that include
\textit{screening mechanisms} that act to hide any modifications in dense
environments. One popular theory is \textit{chameleon
gravity}\cite{Khoury:2003aq,Khoury:2003rn}. This theory includes an extra
scalar degree of freedom coupled to matter, which
results in a new or \textit{fifth-} force. This would usually violate laboratory
bounds but the chameleon mechanism acts to
increase the field's mass by several orders of magnitude in dense environments,
rendering the range of the force shorter than
current experiments can probe \cite{Kapner:2006si}. The force-profile for a
spherical object is shown in figure
\ref{fig:screening}. 

There is a \textit{screening radius} $\rs$, inside which the total force is
simply the Newtonian one but
outside which the value of Newton's constant $G$ is increased. When this is zero
the object is fully unscreened and when this is
close to the object's radius it is said to be screened. The theory is
parametrised by two model-independent quantities, $\chi_0$
and $\alpha$. Outside the screening radius, the effective value of $G$ is 
\begin{equation}
 G\eff(r)= G\left[1+\alpha\left(1-\frac{M(r)}{M(\rs)}\right)\right]\quad r>\rs,
\end{equation}
where $M(r)$ is the mass enclosed by a sphere of radius $r$. $\alpha$ then
parametrises the strength of the fifth-force relative
to the Newtonian one and is
typically $\mathcal{O}(1)$. For example, $f(R)$ theories are chameleon models
with $\alpha=1/3$ \cite{Brax:2008hh}. $\chi_0$, the
\textit{self-screening} parameter, parametrises how efficient an object is at
screening itself. An object of radius $R$ with
Newtonian potential $\pn\equiv GM/R$ is screened if $\chi_0<\pn$, otherwise it
is at least partially
unscreened.
Objects can additionally be screened by the Newtonian potential of their
environment. The Milky Way is necessarily screened,
which imposes the constraint\footnote{In fact, this constraint is not
experimentally confirmed. If the Milky Way is screened by the local group it may
be relaxed to $\chi_0\le10^{-4}$ which has been placed using cluster statistics
\cite{Schmidt:2008tn}. } $\chi_0\le\pn^{\textrm{Milky Way}}\sim10^{-6}$. 
 This means that in order to probe smaller
values one must look at more under-dense objects, which are post-main-sequence
stars and dwarf galaxies with potentials of
$\mathcal{O}(10^{-7})$ and $\mathcal{O}(10^{-8})$ respectively.

In this essay we will examine the modified behaviour of stars in these theories
and use them to place the strongest constraints
to date. Cepheid variable stars pulsate faster at fixed luminosity owing to the
increased strength of gravity. The period-luminosity relation used to measure distances to other galaxies is
calibrated on local
group stars and is hence incorrect in unscreened dwarf galaxies. By comparing
the estimated Cepheid distances to a sample of unscreened galaxies with those
found using a different method that is insensitive to gravitational physics we
can place new bounds on
the model parameters. Until recently, this was not possible but
\cite{Cabre:2012tq} have used data from the Sloan Digital Sky
Survey ({\scriptsize SDSS}) to provide a map of unscreened galaxies in the
nearby universe.

\section{Stars in Modified Gravity}

Stars are spheres of gas that burn nuclear fuel at their centres to provide an
outward pressure gradient to combat gravitational
collapse. In unscreened stars, the gravitational force is stronger and one would
therefore expect the rate of nuclear burning to
be larger. As a consequence, an unscreened star of fixed mass is brighter,
hotter and more ephemeral than its screened
counterpart. In \cite{Davis:2011qf} we presented a semi-analytic model of
main-sequence stars in chameleon gravity and found that
this is indeed the case. The important physics is captured by the hydrostatic
equilibrium equation
\begin{equation}
 \frac{\dd P}{\dd r} = -\frac{GM(r)\rho(r)}{r^2},
\end{equation}
which must be satisfied if the gravitational force is to be balanced by the
pressure to maintain equilibrium. All of the
gravitational physics is encoded in this equation and since chameleon gravity
does not alter any other physics, one simply needs
to replace $G$ by $G\eff(r)$ in order to solve for the modified stellar
properties. 

\begin{figure}[t,h]
\centering
\includegraphics[width=0.95\textwidth]{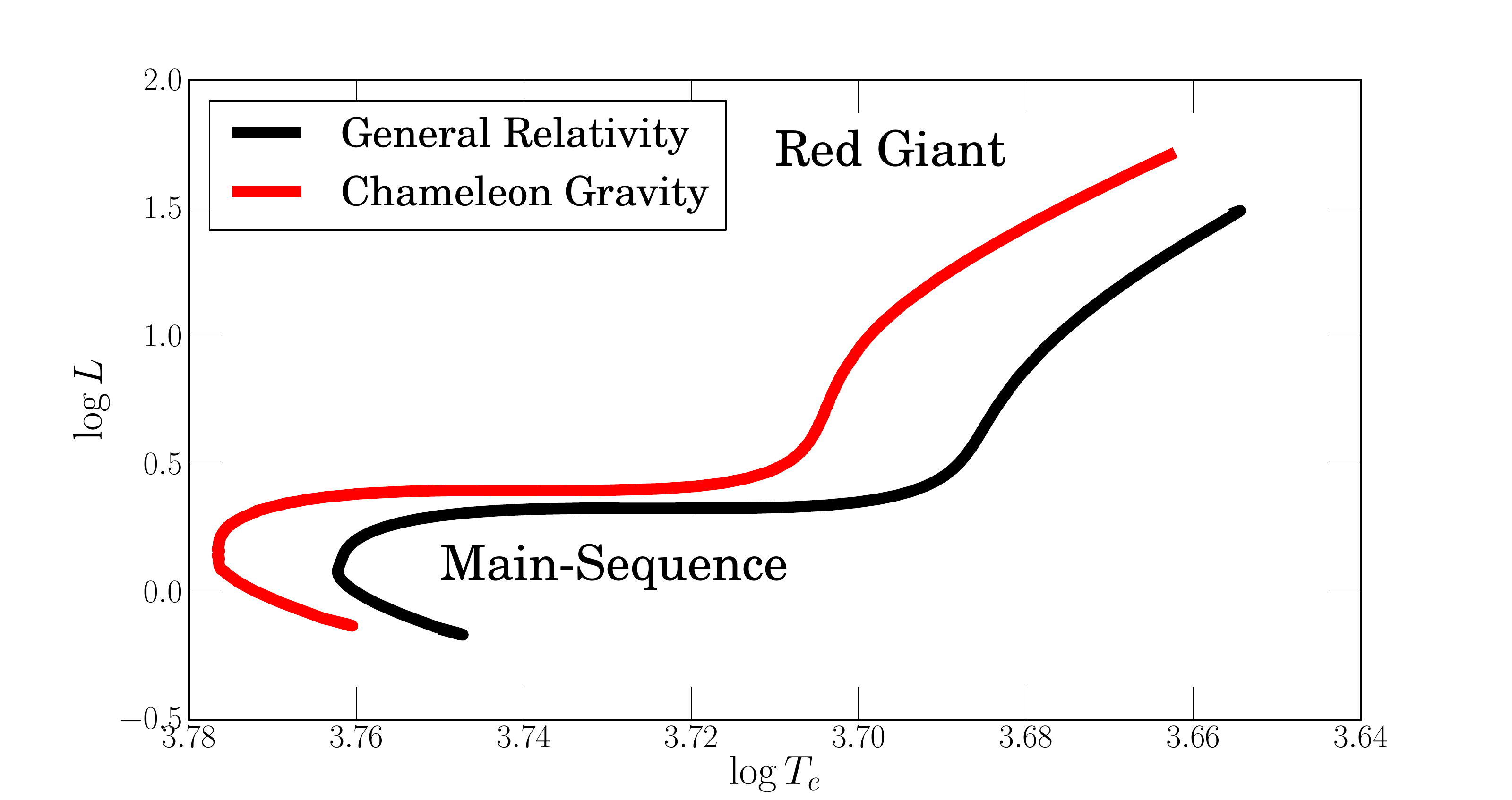}
\caption{The evolution of a 1 $M_\odot$ star in the Hertzprung-Russell diagram
when the theory of gravity is general relativity
(black line) and chameleon gravity (red line) with $\chi_0=10^{-6}$ and
$\alpha=1/3$ corresponding to $f(R)$ gravity.} 
\label{fig:mesa}
\end{figure}

We have updated the stellar structure code {\scriptsize MESA}
\cite{Paxton:2010ji,Paxton:2013pj} to include this modified equation
and the resultant predictions are accurate enough to allow a comparison with
observational data. As an example, the
Hertzsprung–-Russell diagram for a solar mass star in general relativity and
$f(R)$ gravity with $\chi_0=10^{-6}$ is shown in
figure \ref{fig:mesa}.

Cepheid variable stars are $5$--$10$ $M_\odot$ stars that have gone off the
main-sequence. When situated in a narrow temperature
gap in the Hertzsprung-–Russell diagram known as the \textit{instability strip}
they pulsate with a known period-luminosity
relation:
\begin{equation}
 M_{\rm V} = a\log\tau + b(B-V) + c,
 \label{eq:PLC}
 \end{equation}
where $a\approx-3$, $b$ and $c$ are constants, $\tau$ is the period of
oscillation and $M_{\rm V}\propto\log d$ --- $d$ is the
distance to the star --- is the V-band magnitude. General relativity predicts
that $\tau\propto G^{-1/2}$ and so in
unscreened stars we expect the period to be shorter. This relation has been
empirically calibrated using local group stars and
hence applies in
screened situations. In unscreened galaxies, the relation will have different
values for the constants and so if one uses this
relation to estimate the distance then it will not be correct and will disagree
with a different distance estimate obtained using
a distance indicator that is insensitive to gravitational physics. Indeed,
perturbing the relation one finds
\begin{equation}
 \frac{\Delta d}{d}\approx -0.3\frac{\Delta G}{G},
\end{equation}
where $\Delta d$ is the difference between the chameleon and general relativity
distance. Since $\Delta G/G$ is
radially-dependent in chameleon theories, we need some procedure for averaging
$G\eff$ over the star. In 1950, Epstein
\cite{1950ApJ...112....6E} numerically calculated a weighting function for the
importance of different regions of the star for the
pulsation and we have recreated this function using the numerical values in his
paper. Using {\scriptsize MESA} models for stars
at the blue edge of the instability strip, we have calculated $\Delta G\equiv
\langle G\eff\rangle-G$ by averaging the radial
profile of $G\eff$ for many different values of $\chi_0$ and $\alpha$ and
calculated the resultant theoretical prediction for
$\Delta d/d$ \cite{Jain:2012tn}.

Tip of the red giant branch (TRGB) distances are found using a method
insensitive to gravitational physics. Any deviation between
the TRGB distance and Cepheid distance therefore probes chameleon gravity.
Figure \ref{fig:dvd} below shows a comparison of the
two distance estimates for a sample of screened and unscreened galaxies taken
from the screening map of \cite{Cabre:2012tq}. One
can see that there is a good agreement and, as a further example, we plot the
best fitting curves for $\Delta d/d$ vs the TRGB
distance and indicate the predictions from two different chameleon models
indicated in the caption. Again, there is a good
agreement with the general relativity prediction $\Delta d/d=0$.

\begin{figure}[t,h]
\centering
\includegraphics[width=0.95\textwidth]{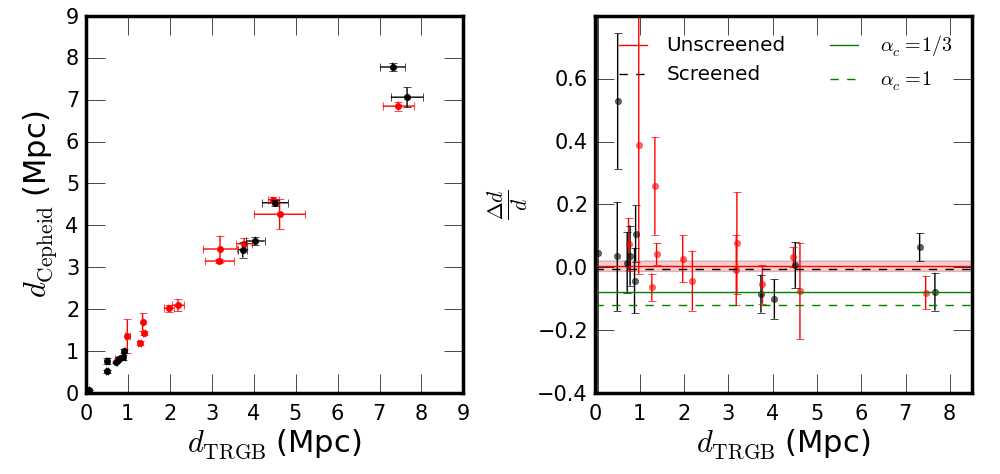}
\caption{\textit{Left}: A comparison of the Cepheid and TRGB distances to the
galaxies in the screened (black data points) and
unscreened (red data points) galaxy samples. \textit{Right}: $\Delta d/d$ as a
function of $d_{\rm TRGB}$. The black and red lines
show
the best-fitting relations for the screened and unscreened galaxy samples
respectively. The solid green line shows our
theoretical prediction for $\chi_0=10^{-6}$, $\alpha=1/3$ and the dashed green
line shows the prediction
for $\chi_0=4 \times 10^{-7}$, $\alpha=1$.} 
\label{fig:dvd}
\end{figure}

\begin{figure}[t,h]
\centering
\includegraphics[width=10cm]{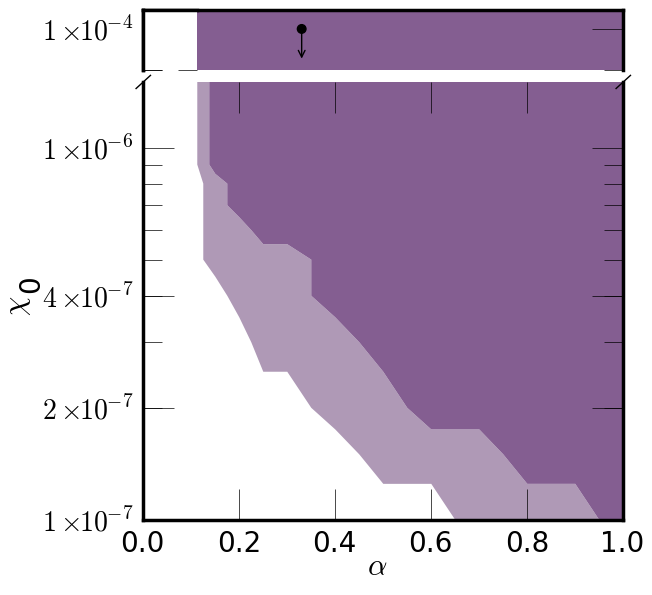}
\caption{The new constraints in the $\chi_0$--$\alpha$ plane. The light and dark
shaded regions show the
regions excluded with 68\% and 95\% confidence respectively. The black arrow
indicates the previous constraint coming from
galaxy cluster statistics.} 
\label{fig:cha}
\end{figure}

We have performed a reduced $\chi^2$ analysis using our predictions for $\Delta
d/d$ and have been able to rule out new
regions of parameter space, three orders of magnitude below those previously
probed \cite{Schmidt:2008tn}. These
constraints are shown in figure \ref{fig:cha}. In particular, we can place the
new bound $\chi_0<4\times10^{-7}$ for $f(R)$
theories. Given these bounds, the only objects in the universe that can be
unscreened are isolated gas clouds, the smallest dwarf galaxies and very massive
($\ge
10M_\odot$) post-main-sequence stars.

\section{Conclusions}

By exploiting the novel features exhibited by stars in chameleon theories of
gravity we have been able to place new
constraints on the model parameters. We have turned the distance ladder, used to
calibrate Supernovae distances that led to the discovery of cosmic acceleration,
into a new test of gravity. 
Our analysis above was data-limited --- we only have 25 galaxies in our
unscreened sample. This is the reason for the jaggedness of the contours in
figure \ref{fig:cha}. With new data, these
constraints can be significantly improved \cite{Sakstein:2013pda}. 

Astrophysical tests of modified gravity do not require dedicated experiments and
can piggyback
on current and upcoming experimental surveys. {\scriptsize SDSS-IV MaNGA} will
soon provide a larger sample of galaxies
and {\scriptsize LSST} will provide data pertaining to variable stars in a
variety of different environments. {\scriptsize
Spitzer} will provide infra-red data that can reduce the errors on the distance
measurements and allowing further constraints to be placed. In addition to
chameleon theories, others based on Vainshtein screening may also be tested by
comparing compact objects with stars and gas.

The constraints we have placed here are the strongest to date and leave only a
small viable region of parameter space. When
$\chi_0<10^{-8}$ there are no unscreened objects in the universe and the
chameleon theories are eliminated for all practical purposes. The
next decade will see a plethora of surveys that can be used to test modified
gravity theories using methods similar to those presented here.

\section{Outlook}

In this essay we have used post-main-sequence stars to place the tightest constraints on chameleon theories to date, but there
are many other modified theories of gravity---and other screening mechanisms---that are the study of modern research (see
\cite{Joyce:2014kja} for a review). With this in mind, one may wonder about the prospect of using other astrophysical systems to
constrain alternate theories of gravity. The fact that similar measurements---TRGB and Cepheid distances---are affected in a
qualitatively different manner when the theory of gravity is different from GR was paramount in allowing us to place new
constraints. The first step towards constraining other theories is then to identify astrophysical systems where different
components respond differently to modified gravity.

One can envision many such systems and so here we will give only one as an illustrative example. Recently, a Cepheid has been
observed in an eclipsing binary \cite{Pietrzynski:2010tr,Marconi:2013tta}. The eclipsing binary method allows one to measure the
masses of both objects (which are both found to be of order $4M_\odot$) but the fact that one is a pulsating Cepheid allows a
second, independent measurement of its mass using the period-luminosity-mass (P-L-M) relation. Both measurements agree to within
1\% and so this system has the potential to place constraints on any theory of gravity where the masses inferred using the P-L-M
relation and the eclipsing binary technique do not agree. These are any theories where either the interior of the Cepheid is
unscreened or the orbital motion of two bodies is not Keplarian, for example, this is the case for theories that screen using
the Vainshtein mechanism \cite{Hiramatsu:2012xj}. To date, this system has not been utilised to constrain alternate theories of
gravity but clearly it merits further investigation.

\section*{Acknowledgements}

JS is grateful to collaborators on the various papers used as the basis for
this essay: Anne-Christine Davis, Eugene A.
Lim and Douglas J. Shaw. We would like to thank Phillip Chang, Lam Hui, Justin
Khoury and Mark Trodden for
insightful discussions. This essay was written whilst JS was a visiting fellow
at the Perimeter Institute for Theoretical Physics.
Research at Perimeter Institute is supported by the Government of Canada through
Industry Canada and by the Province of Ontario
through the Ministry of Economic Development \& Innovation.

\bibliographystyle{ws-ijmpd}
\bibliography{ref}
\end{document}